\documentclass[aps,pra,showpacs,twocolumn,superscriptaddress,tightenlines]{revtex4}

\usepackage{amssymb}
\usepackage{amsmath}
\usepackage{graphicx}
\usepackage{mathrsfs}
\usepackage{color}
\usepackage[colorlinks=true,linkcolor=blue,citecolor=blue]{hyperref}

\begin{document}

\title
{Tunneling-induced transparency and Autler-Townes splitting in a triple quantum dot}

\author{Xiao-Qing Luo}
\affiliation{Quantum Physics and Quantum Information Division, Beijing Computational Science Research Center, Beijing 100094,
China}
\author{Zeng-Zhao Li}
\affiliation{Quantum Physics and Quantum Information Division, Beijing Computational Science Research Center, Beijing 100094,
China}
\author{Jun Jing}
\affiliation{Institute of Atomic and Molecule Physics, Jilin University, Changchun 130012, China}
\affiliation{Quantum Physics and Quantum Information Division, Beijing Computational Science Research Center, Beijing 100094,
China}
\author{Tie-Fu Li}
\affiliation{Institute of Microelectronics, Department of Micro and Nanoelectronics and Tsinghua National Laboratory of Information Science and Technology, Tsinghua University, Beijing 100084, China}
\affiliation{Quantum Physics and Quantum Information Division, Beijing Computational Science Research Center, Beijing 100094,
China}
\author{Ting Yu}
\affiliation{Department of Physics and Engineering Physics, Center for Controlled Quantum Systems, Stevens Institute of Technology,
Hoboken, New Jersey 07030, USA}
\affiliation{Quantum Physics and Quantum Information Division, Beijing Computational Science Research Center, Beijing 100094,
China}
\author{J. Q. You}
\affiliation{Quantum Physics and Quantum Information Division, Beijing Computational Science Research Center, Beijing 100094,
China}

\date{\today}

\begin{abstract}
We theoretically investigate the tunneling-induced transparency (TIT) and the Autler-Townes (AT) doublet and triplet in a triple-quantum-dot system. For the resonant tunneling case, we show that the TIT induces a transparency dip in a weak-tunneling regime and no anticrossing occurs in the eigenenergies of the system Hamiltonian. However, in a strong-tunneling regime, we show that the TIT evolves to the AT splitting, which results in a well-resolved doublet and double anticrossings. For the off-resonance case, we demonstrate that, in the weak-tunneling regime, the double TIT is realized with a new detuning-dependent dip, where the anticrossing is also absent. In the strong-tunneling regime, the AT triplet is realized with triple anticrossings and a wide detuning-dependent transparency window by manipulating one of the energy-level detunings. Our results can be applied to quantum measurement and quantum-optics devices in solid systems.
\end{abstract}

%42.50.Gy	Effects of atomic coherence on propagation, absorption, and amplification of light;
           %electromagnetically induced transparency and absorption
%78.67.Hc	Quantum dots
\maketitle

\section{Introduction}

Quantum coherence and interference effects can lead to considerably interesting phenomena of quantum optics such as lasing without inversion~\cite{HSZ}, coherent population trapping~\cite{Arimondo1996}, correlated spontaneous emission~\cite{scully1985}, and electromagnetically induced transparency (EIT)~\cite{Harris1997,FIM2005,Marangos1998,HHDB1999,LDBH2001,PFMWL2001,HI1996,HH1999}. As a phenomenon closely related to EIT, Autler-Townes (AT) splitting~\cite{AT1955,Cohen-Tannoudji1996} is indicated by a \emph{level anticrossing} in the energy spectrum and a transparency window owing to the AT doublet rather than the quantum interference. This phenomenon has been utilized to measure the state of the electromagnetic field~\cite{Zubairy1996,HSZ1997,IZ2002}, as well as the AT triplet and multiplet spectroscopy~\cite{GQZZ2007,Ghafoor2014}. Both EIT and AT splitting have been investigated theoretically and experimentally in different quantum systems, including atomic and molecular systems~\cite{LFZR1997,HHDB1999,MFBH2010,ZTH2013}, solid-state and metamaterials systems~\cite{LFSM2005,ZGWL2007,PFZP2008,LLWK2009}, superconducting quantum circuits~\cite{AMN2002,AAZP2010,ADS2011,YOU2011,XAYN2013,SLIY2014} and whispering-gallery-mode optical resonators~\cite{XSPS2006,TKT2007,LXJL2012,POCN2014}. It is also interesting to investigate EIT and AT splitting in semiconductor nanostructures because the trapped carriers behave like atoms and can be conveniently manipulated via external fields.

In the semiconductor quantum dots (QDs), excitons form bound states and play an important role in the optical properties of these systems~\cite{Knox1963,SEZM2005,MFBW2007,XSBS2007,BABW1994,SMEB2007,JRHF2008,GBDK2009,LWSG2003,ALPB2006}. Moreover, tunneling-induced transparency (TIT)~\cite{YZ2006,BPD2008,BSVD2012,BSVA2013} can occur for the excitonic states, which is similar to EIT in a three-level atomic system, but no pump field is needed to apply to the excitonic system. As shown in Refs.~\cite{BSVD2012,BSVA2013}, there is an evidence regarding the coexistence of both TIT and AT splitting in the intermediate regime, when the tunneling coupling is slightly above a threshold in a double QD system. However, a triple QD system can offer new possibilities to study intriguing phenomena that are not observed in single and double QD systems~\cite{HSKH2012,DLoss2003,WBMW1995,GSSZ2006,GJHL2008}. In the present paper, we show that when the electron is resonantly tunneling, which is equivalent to the case considered in a double QD system, the coexistence of both TIT and AT splitting does not occur in the triple QD system and the threshold of the tunneling coupling just corresponds to a transition point. More specifically, we find that in the weak-tunneling regime, the TIT presents a transparency dip and no anticrossing occurs in the eigenenergies of the system Hamiltonian. However, in the strong-tunneling regime, the TIT evolves to the AT splitting exhibiting \emph{double anticrossings} and a well-resolved doublet. For the off-resonance case, the double TIT can be realized, with a detuning-dependent TIT dip in the weak-tunneling regime. Moreover, the \emph{triple anticrossings} in the strong-tunneling regime reveal AT triplet, accompanied by a wide red-shifted (blue-shifted) transparency window in the presence of a blue (red) detuning.

The paper is organized as follows. In Sec.~\ref{s2}, we present a theoretical model for the effective four-level system realized in a triple QD. In Sec.~\ref{s3}, we demonstrate the difference between the TIT and the AT doublet via resonant electron tunneling in the weak- and strong-tunneling regimes, respectively. In Sec.~\ref{s4}, we show the double TIT with a detuning-dependent TIT dip. The AT triplet reveals triple anticrossings and a wide detuning-dependent transparency window. Finally, conclusions are given in Sec.~\ref{s5}.

\section{Triple quantum-dot system}\label{s2}

\begin{figure}[htbp]
\centering
\includegraphics[width=0.46\textwidth]{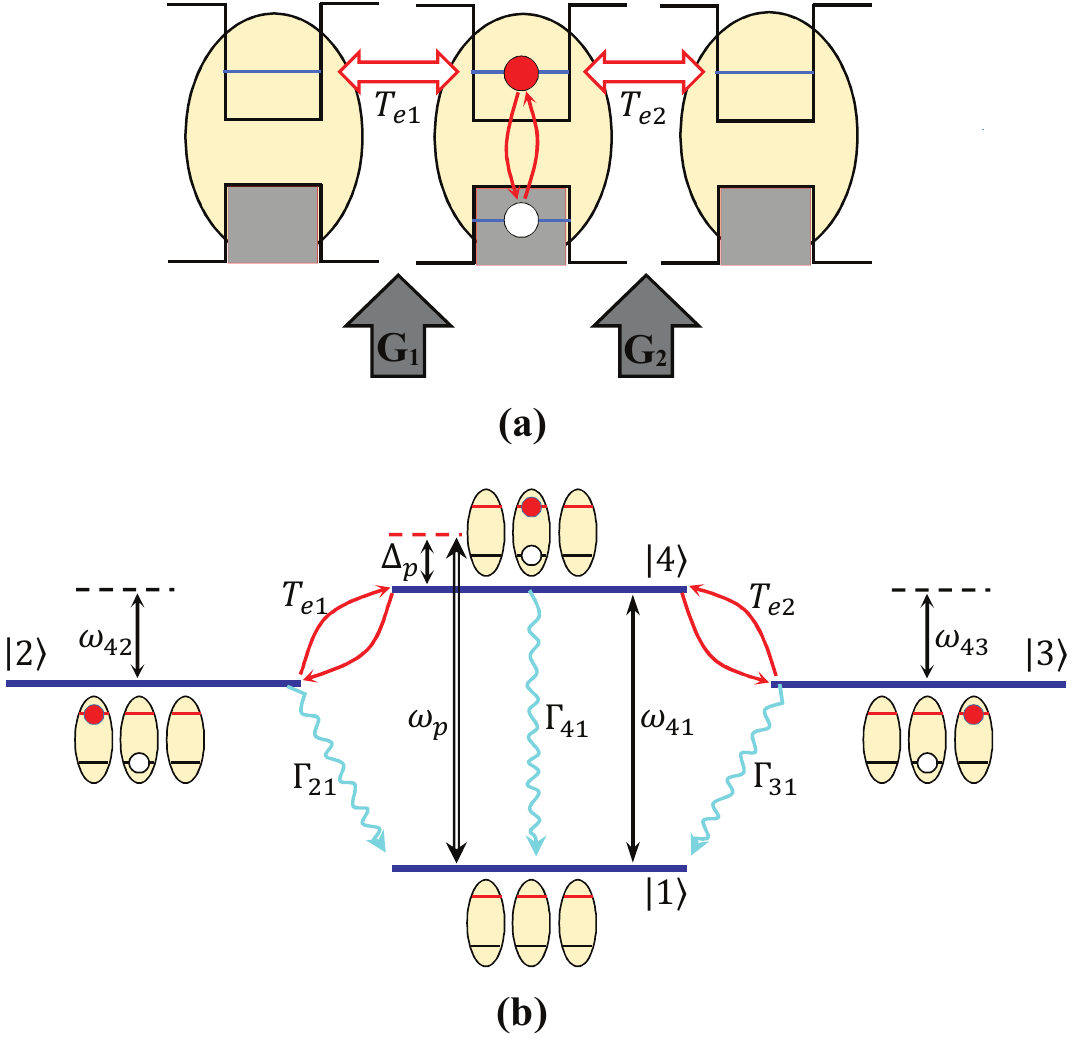}
\caption{(Color online) (a) Schematic energy-level diagram of a triple QD system. (b) Excitation scheme of the triple QD system, as determined by the Rabi frequency $\Omega_p$ (which is proportional to the probe-field strength), decoherence channels $\Gamma_{i1}$ ($i$=2, 3, 4), energy-level difference $\omega_{4j}$ ($j$=1, 2, 3), probe-field detuning $\Delta_p=\omega_p-\omega_{41}$ from the energy-level difference $\omega_{41}$, and tunneling coupling $T_{ei}$ ($i$=1, 2). Driven by a pulsed laser field, one electron can be excited from the valence band to the conduction band to form a direct exciton state $|4\rangle$ inside the central QD.  The electric field allows the electron to tunnel from the central dot to the left (right) dot to form an indirect exciton states $|2\rangle$ ($|3\rangle$). Here $|1\rangle$ denotes the state with no exciton inside this triple QD.} \label{energylevel}  
\end{figure}

We study a triple-QD molecule consisting of three aligned QDs separated by two barriers, which inhibits hole tunneling between valance bands [see Fig.~\ref{energylevel}(a)]. This QD system can be achieved using, e.g., a self-assembled (In, Ga) As triple QD fabricated on a GaAs (001) substrate by molecule beam epitaxy or in-situ atomic layer precise etching, corresponding to a homogeneous triple-QD along the [1\={1}0] direction~\cite{HSKH2012}. The system is driven by a weak probe laser field with frequency $\omega_p$ and the Rabi frequency $\Omega_p$ corresponds to the driving strength for generating the direct exciton state in the central dot. As shown in Fig. \ref{energylevel}(b), we assume that an electron can be excited from the valence band to the conduction band via a pulsed laser field to form a direct exciton state (denoted as $|4\rangle$) only in the central QD, where $|1\rangle$ denotes the state without any exciton in the triple QD in the absence of an optical pulse. The gate voltages, denoted by $\mathrm{G_1}$ and $\mathrm{G_2}$ in Fig. \ref{energylevel}(a), allow the electron to tunnel from the central dot to either the left or right dot, yielding an indirect exciton state $|2\rangle$ or $|3\rangle$.  The Hamiltonian for this triple-QD system reads (setting $\hbar=1$)
\begin{eqnarray}
H=\sum_{i=1}^4\omega_i\sigma_{ii}-(\Omega_{p}e^{-i\omega_p t}\sigma_{41}+T_{e1}\sigma_{42}+T_{e2}\sigma_{43}+\text{H.c.}),
\end{eqnarray}
where $\sigma_{ij}\equiv |i\rangle\langle j|$, and $\Omega_{p}=\mu_{14}\mathcal{E}_p/2\hbar$ is the Rabi frequency of the probe laser field, with $\mu_{14}$ being the electric-dipole momentum matrix element between $|1\rangle$ and $|4\rangle$, and $\mathcal{E}_p$ the electric-field amplitude of the probe field. $T_{e1}$ ($T_{e2}$) denotes the tunneling couplings between the central dot and the left (right) dot.

By using the unitary transformation $U(t)=\exp[-i\omega_p(\sum_{i=2}^{4}\sigma_{ii})t]$ to remove the time-dependent oscillatory terms~\cite{XAYN2013}, the Hamiltonian in this interaction picture can be written as
\begin{eqnarray}
H_{I}&=&-\Delta_{p}\sigma_{44}-\Delta_{2}\sigma_{22}-\Delta_{3}\sigma_{33} \notag \\ &&-(\Omega_{p}\sigma_{41}+T_{e1}\sigma_{42}+T_{e2}\sigma_{43}+\text{H.c.}), \label{H}
\end{eqnarray}
with $\Delta_p=\omega_p-\omega_{41}$, $\Delta_2=\Delta_p+\omega_{42}$, and $\Delta_3=\Delta_p+\omega_{43}$. Here $\Delta_p$ denotes the detuning of the probe field from $\omega_{41}$, and $\omega_{ij}$ is the energy difference between $|i\rangle$ and $|j\rangle$.

The dynamics of the system can be described by a Lindblad master equation:
\begin{eqnarray}
\partial_t\rho=-\frac{i}{\hbar}[H_{I},\rho]+\sum_{i=2}^4\Big(\frac{\Gamma_{i1}}{2}\mathcal{D}[\sigma_{1i}]\rho+\gamma_{i}^{\phi}\mathcal{D}[\sigma_{ii}]\rho\Big),
\end{eqnarray}
where $\mathcal{D}[\hat{\mathcal{O}}]\rho=2\hat{\mathcal{O}}\rho \hat{\mathcal{O}}^\dag-\hat{\mathcal{O}}^\dag \hat{\mathcal{O}} \rho-\rho \hat{\mathcal{O}}^\dag \hat{\mathcal{O}}$, $\Gamma_{i1}$ are the relaxation rates between $|i\rangle$ and $|1\rangle$, and $\gamma_{i}^{\phi}$ describe the pure dephasing rates of the states $|i\rangle$ ($i$=2, 3, 4). The decoherence of the excitonic states is induced by both relaxation and pure dephasing processes. Explicitly, the coupled differential equations for the density matrix $\rho_{ij}$ elements are given as follows:
\begin{eqnarray}
\partial_t\rho_{11}&=&\Gamma_{21}\rho_{22}+\Gamma_{31}\rho_{33}
+\Gamma_{41}\rho_{44}-i\Omega_{p}\rho_{14}+i\Omega_{p}^{*}\rho_{41},  \notag\\
\partial_t\rho_{22}&=&-\Gamma_{21}\rho_{22}-i T_{e1}\rho_{24}+i T_{e1}\rho_{42},  \notag\\
\partial_t\rho_{33}&=&-\Gamma_{31}\rho_{33}-i T_{e2}\rho_{34}+i T_{e2}\rho_{43},    \notag\\
\partial_t\rho_{44}&=&-\Gamma_{41}\rho_{44}+i\Omega_{p}\rho_{14}+i T_{e1}\rho_{24}+i T_{e2}\rho_{34}  \notag\\
&&-i\Omega_{p}^{*}\rho_{41}-i T_{e1}\rho_{42}-i T_{e2}\rho_{43},   \notag\\
\partial_t\rho_{12}&=&i(\Delta_2+i\Gamma_{2})\rho_{12}-i T_{e1}\rho_{14}+i\Omega_{p}^{*}\rho_{42},      \notag\\
\partial_t\rho_{13}&=&i(\Delta_3+i\Gamma_{3})\rho_{13}-i T_{e2}\rho_{14}+i\Omega_{p}^{*}\rho_{43},         \notag\\
\partial_t\rho_{14}&=&i(\Delta_p+i\Gamma_{4})\rho_{14}-i T_{e1}\rho_{12}-i T_{e2}\rho_{13} \notag \\
&&-i\Omega_{p}^{*}(\rho_{11}-\rho_{44}),     \notag\\
\partial_t\rho_{23}&=&i(\Delta_3-\Delta_2+i\gamma_{23})\rho_{23}-i T_{e2}\rho_{24}+i T_{e1}\rho_{43},      \notag\\
\partial_t\rho_{24}&=&i(\Delta_p-\Delta_2+i\gamma_{24})\rho_{24}
-i\Omega_{p}^{*}\rho_{21}-i T_{e2}\rho_{23}  \notag\\
&&-i T_{e1}(\rho_{22}-\rho_{44}),  \notag\\
\partial_t\rho_{34}&=&i(\Delta_p-\Delta_3+i\gamma_{34})\rho_{34}
-i\Omega_{p}^{*}\rho_{31}-i T_{e1}\rho_{32} \notag\\
&&-i T_{e2}(\rho_{33}-\rho_{44}), \label{DME}
\end{eqnarray}
with $\Gamma_{i}=\Gamma_{i1}/2+\gamma_{i}^{\phi}$ ($i$=2, 3 and 4), and
$\gamma_{ij}=(\Gamma_{i1}+\Gamma_{j1})/2+\gamma_{i}^{\phi}+\gamma_{j}^{\phi}$ ($i=2, 3$; $j=3, 4$).

\section{Tunneling-induced transparency and Autler-Townes doublet}\label{s3}

\begin{figure}[htbp]
\begin{center}
\includegraphics[width=0.48\textwidth]{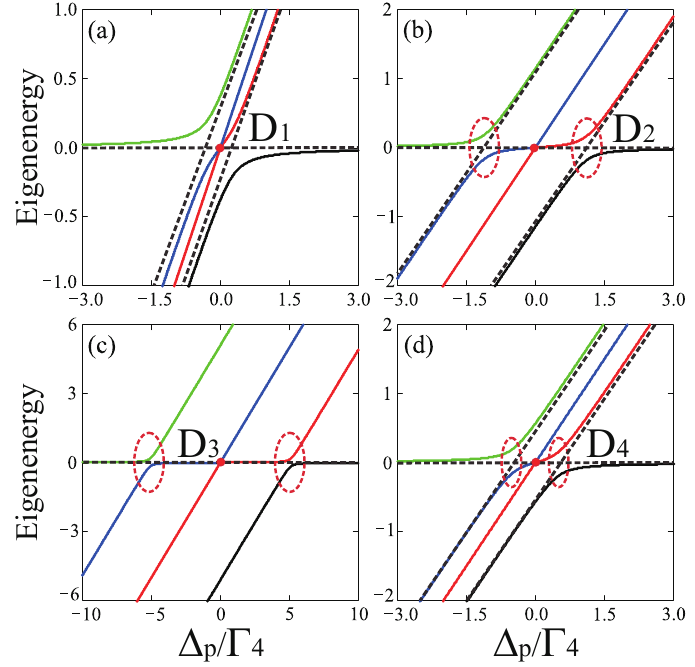}
\caption{(Color online) Eigenenergies of the Hamiltonian (\ref{H}) as a function of the probe-field detuning $\Delta_p$ at different values of the tunneling coupling strength: (a) $T_{e1}=T_{e2}=\Gamma_4/5$, (b) $T_{e1}=\Gamma_4/2, T_{e2}=\Gamma_4$, (c) $T_{e1}=\Gamma_4, T_{e2}=5\Gamma_4$, and (d) $T_{e1}=\Gamma_4/2, T_{e2}=10^{-1}\Gamma_4$, where $\Gamma_4=10\mu eV$. Other parameters are chosen as $\Gamma_2=\Gamma_3=10^{-3}\Gamma_4$, $\Omega_p=10^{-2}\Gamma_4$, and $\omega_{42}=\omega_{43}=0$. Here $D_i$ ($i$=1, 2, 3, 4) labels the degenerate points in the eigenenergy diagrams and the red dashed loops are used to highlight the anticrossing points.} \label{eigenenergies1}
\end{center}
\end{figure}

\begin{figure*}[htbp]
\begin{center}
\includegraphics[scale=0.9]{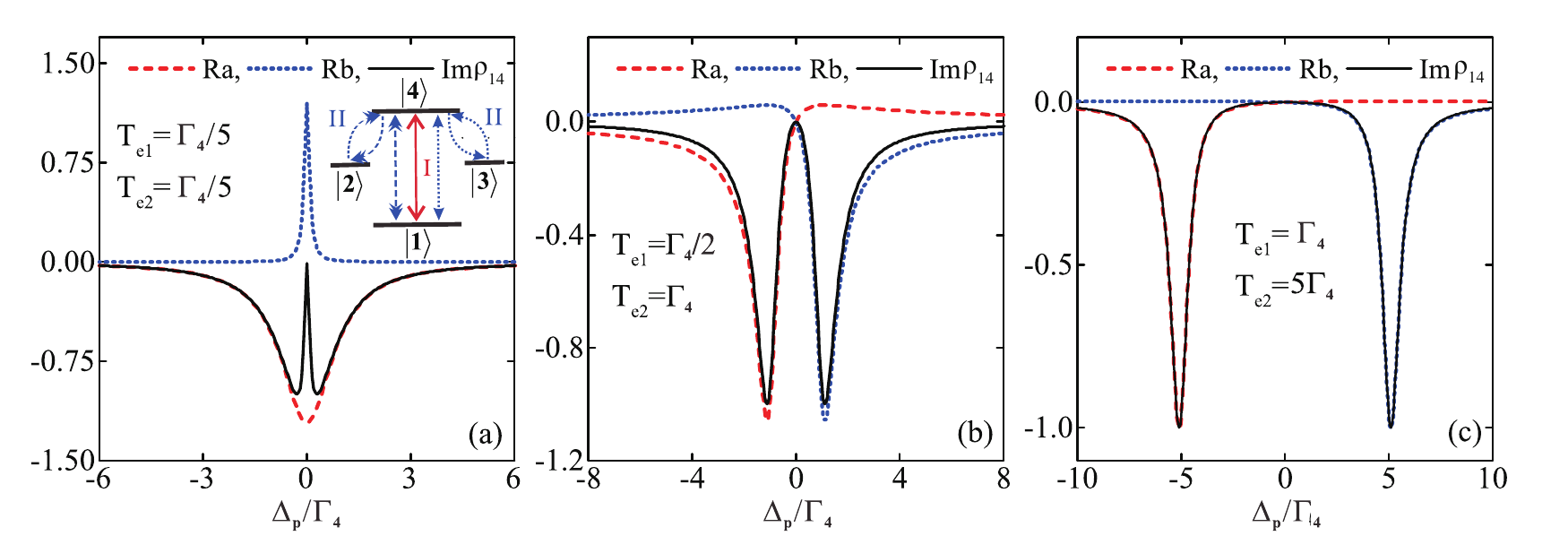}
\caption{(Color online) $\mathrm{Im}(\rho_{14})$ (black curves) and the imaginary part of the two resonances $R_a$ (red dashed curves) and $R_b$ (blue dotted curves) as a function of the probe detuning $\Delta_p$ at different values of the tunneling coupling strength: (a) $T_{e1}=T_{e2}=\Gamma_4/5$, (b) $T_{e1}=\Gamma_4/2, T_{e2}=\Gamma_4$, and (c) $T_{e1}=\Gamma_4, T_{e2}=5\Gamma_4$. Other parameters are the same as in Fig. \ref{eigenenergies1}.} \label{SR}
\end{center}
\end{figure*}
From Eq.~(\ref{DME}), when the time is explicitly shown in the equations of $\partial_t\rho_{12}$, $\partial_t\rho_{13}$, and $\partial_t\rho_{14}$, it follows that 
\begin{equation}
\begin{aligned}
\partial_t\rho_{12}(t)=&i(\Delta_2+i\Gamma_{2})\rho_{12}(t)-i T_{e1}\rho_{14}(t)+i\Omega_{p}^{*}\rho_{42}(t),      \\
\partial_t\rho_{13}(t)=&i(\Delta_3+i\Gamma_{3})\rho_{13}(t)-i T_{e2}\rho_{14}(t)+i\Omega_{p}^{*}\rho_{43}(t),       \\
\partial_t\rho_{14}(t)=&i(\Delta_p+i\Gamma_{4})\rho_{14}(t)-i T_{e1}\rho_{12}-i T_{e2}\rho_{13}(t)  \\
&-i\Omega_{p}^{*}(\rho_{11}(t)-\rho_{44}(t)).     \\  
\end{aligned}
\label{BSQ}
\end{equation}
The absorption and dispersion coefficients are respectively proportional to the imaginary and real parts of the density matrix element $\rho_{14}$ in the steady state~\cite{YZ2006,AHDN2010,BSVD2012,BSVA2013}. With a weak probe field ($\Omega_p\ll T_{e1}, T_{e2}$) acting on the considered triple QD system, the term $\Omega_p^{*}\rho_{42}(t)$ and $\Omega_p^{*}\rho_{43}(t)$  in Eq.~(\ref{BSQ}) can be respectively approximated by  $\Omega_p^{*}\rho_{42}(0)$ and $\Omega_p^{*}\rho_{43}(0)$, and the term $\Omega_p^{*} [\rho_{11}(t)-\rho_{44}(t)]$ can also be approximated by $\Omega_p^{*} [\rho_{11}(0)-\rho_{44}(0)]$. Moreover, the system is assumed to be initially in the ground state $|1\rangle$, so $\rho_{11}(0)=1$ and $\rho_{42}(0)=\rho_{43}(0)=\rho_{44}(0)=0$. Then, Eq.~(\ref{BSQ}) becomes
\begin{equation}
\begin{aligned}
\partial_t\rho_{12}&=i(\Delta_2+i\Gamma_{2})\rho_{12}-i T_{e1}\rho_{14},      \\
\partial_t\rho_{13}&=i(\Delta_3+i\Gamma_{3})\rho_{13}-i T_{e2}\rho_{14},       \\
\partial_t\rho_{14}&=i(\Delta_p+i\Gamma_{4})\rho_{14}-i T_{e1}\rho_{12}-i T_{e2}\rho_{13}-i\Omega_{p}^{*}.    \\  
\end{aligned}
\end{equation}
In the steady-state, $\partial_t\rho_{12}=\partial_t\rho_{13}=\partial_t\rho_{14}=0$, so we have
\begin{equation}
\rho_{14}=\frac{d_{2}d_{3}\Omega_p^{*}}{d_{2}d_{3}d_{4}-T_{e1}^2d_{3}-T_{e2}^2d_{2}},\label{rho14}
\end{equation}
where $d_{2(3)}=\Delta_{2(3)}+i\Gamma_{2(3)}$, and $d_4=\Delta_{p}+i\Gamma_{4}$. It can be seen that when $T_{e2}=0$, i.e., in the absence of the right-side electron tunneling in Fig.~\ref{energylevel}, Eq.~(\ref{rho14}) reduces to the result for the linear response of a $\Lambda$-type three-level QD system~\cite{YZ2006,BSVD2012}. For the triple-QD system described by Hamiltonian (\ref{H}), when $\Delta_p=0$, degenerate points occur in the eigenenergy spectrum [see the red points $D_i$ ($i$=1, 2, 3, 4) in Fig.~\ref{eigenenergies1}], where the dressed-state analysis is used under the condition $\omega_{42}=\omega_{43}=0$ with different tunneling-coupling strengths of $T_{e1}$ and $T_{e2}$. Moreover, the degenerate dark states $|\psi_1\rangle_{\mathrm{dark}}$ and $|\psi_2\rangle_{\mathrm{dark}}$ can be analytically obtained as
\begin{subequations}
\begin{align}
&|\psi_1\rangle_{\mathrm{dark}}=c_1[-T_{e2}|1\rangle+\Omega_p|3\rangle],  \\ &|\psi_2\rangle_{\mathrm{dark}}=c_2[\Omega_pT_{e1}|1\rangle
+T_{e1}T_{e2}|3\rangle-(\Omega_p^2+T_{e2}^2)|2\rangle],
\end{align}
\end{subequations}
where $c_1=1/\sqrt{\Omega_p^2+T_{e2}^2}$, and $c_2=1/\sqrt{[\Omega_p^2T_{e1}^2+T_{e1}^2T_{e2}^2+(\Omega_p^2+T_{e2}^2)^2]}$. Note that when $T_{e2}=0$, the dark state $|\psi_2\rangle_{\mathrm{dark}}$ is equivalent to the dark state in a $\Lambda$-type three-level QD system, $|\psi\rangle_{\mathrm{dark}}=[T_{e1}|1\rangle-\Omega_p|2\rangle]/\sqrt{\Omega_p^2+T_{e1}^2}$.

In the case of an electron resonantly tunneling in the triple-QD system, i.e., $\omega_{42}=\omega_{43}=0$, the degenerate dark state only leads to a single transparency window. Physically, it is useful to split  $\rho_{14}$ into two terms $R_a$ and $R_b$, which represent the first resonance ``I'' [see the red solid line with arrows in the inset of Fig.~\ref{SR}(a)] and the second resonance ``II'' [see the blue dashed or dotted lines with arrows in  the inset of Fig.~\ref{SR}(a)], respectively,
\begin{equation}
\rho_{14}=R_a+R_b=\frac{R_+}{\Delta_p-\Delta_+}+\frac{R_-}{\Delta_p-\Delta_-},\label{rho14sr}
\end{equation}
where $\Delta_{\pm}=[-(\omega_{42}+i\Gamma_{2}+i\Gamma_{4})\pm\alpha]/2$ and $R_{\pm}=\pm(\Delta_++\omega_{42}+i\Gamma_{2})/\alpha$, with $\alpha^2=(\omega_{42}+i\Gamma_{2}-i\Gamma_{4})^2+4T_{e1}^2+4d_{2}T_{e2}^2/d_{3}$. By solving $\alpha^2=0$~\cite{POCN2014,SLIY2014,AHDN2010}, a transition point turns out at the threshold coupling strength $T_t\equiv\Gamma_{4}/2$ due to $\omega_{42}=\omega_{43}=0$ and $\Gamma_{2}=\Gamma_{3}$. Thus, the threshold value of $T_t$ separates TIT in the weak-tunneling regime ($0<\sqrt{T_{e1}^2+T_{e2}^2}<T_t$) from AT splitting in the strong-tunneling regime ($\sqrt{T_{e1}^2+T_{e2}^2}>T_t$).

\subsection{The weak-tunneling regime}

In the weak regime with $0<\sqrt{T_{e1}^2+T_{e2}^2}<T_t$, $\alpha$ is a pure imaginary number. It gives rise to a pure real number $R_{\pm}=1/2\mp\varepsilon_1/|\alpha|$, with $\varepsilon_1=(\Gamma_{4}-\Gamma_{2})/2$, and a pure imaginary number $\Delta_{\pm}=i(-\varepsilon_2\pm|\alpha|/2)$, with $\varepsilon_2=(\Gamma_{4}+\Gamma_{2})/2$. Thus, the imaginary part of $\rho_{14}$ is given by
\begin{equation}
\mathrm{Im}(\rho_{14})_{\mathrm{TIT}}=\frac{C_1}{\Delta_p^{2}+\Delta_{+}^2}-\frac{C_2}{\Delta_p^{2}+\Delta_{-}^2},\label{ImTIT}
\end{equation}
where $C_1=(1/2-\varepsilon_1/|\alpha|)(-\varepsilon_2+|\alpha|/2)$, and $C_2=(1/2+\varepsilon_1/|\alpha|)(\varepsilon_2+|\alpha|/2)$.

As shown in Fig.~\ref{SR}(a), when the tunneling couplings $T_{e1}$ and $T_{e2}$ are both weak, the optical absorption profile has two peaks corresponding to the first and second resonances, respectively. The second resonance $R_b$ [see the blue dotted curve in Fig.~\ref{SR}(a)] is narrow and positive. Obviously, this resonence is above the first resonance $R_a$ [see the red dashed curve in Fig.~\ref{SR}(a)], because the latter is negative. The reduction of the optical absorption coefficient in this weak coupling case results from the positive resonance that produces a narrow dip. It is clear that the Lorentzian peak and valley generate a destructive interference that leads to the realization of TIT. In particular, as shown in Fig.~\ref{eigenenergies1}(a), the eigenenergies of this system Hamiltonian do not display an obvious anticrossing. This is in sharp contrast to the case with double anticrossings [see the red dashed loops in Fig.~\ref{eigenenergies1}(d)], when the relative tunneling couplings are slightly stronger than the threshold coupling strength $T_t$. By comparing the profile of the second resonance with the total absorption spectrum [the black solid curve in Fig.~\ref{SR}(a)], the TIT does not change the overall shape of the absorption profile. However, it gives rise to a sudden dip at resonance, i.e., a spectrally narrow transparency window. Physically, the quantum interference becomes significant when the separation between the two peaks in the total absorption profile is less than or even comparable to the decay rate $\Gamma_4$.

In the low-saturation limit~\cite{AHDN2010}, which associates with the pathways to the resonances, the resonances reduce to more readable forms linking the excitation pathways to the pair of resonances in the bare-state picture. It is shown that each resonance represents a microscopic excitation pathway. As shown in the inset of Fig.~\ref{SR}(a), the red solid line represents the first resonance $R_a$, which involves the absorption of a photon to form the direct exciton state $|4\rangle$ by exciting an electron from the valance band to the conduction band. The second resonance $R_b$ [the blue dashed (dotted) lines in the inset of Fig.~\ref{SR}(a)] involves three processes, i.e., a probe photon is absorbed to form the direct exciton state $|4\rangle$, then the electron tunnels from the central dot to either the left or right dot, forming the indirect exciton state $|2\rangle$ or $|3\rangle$, and finally the electron tunnels back to the central dot to recover the direct exciton state $|4\rangle$. Therefore, the cancellation of the negative resonance $R_a$ and the positive resonance $R_b$ in Eq.~(\ref{ImTIT}) is a consequence of the destructive interference involved in the TIT.

\subsection{The strong-tunneling regime}

In the strong-tunneling regime with $T_t<\sqrt{T_{e1}^2+T_{e2}^2}$, $\alpha=2\sqrt{T_{e1}^2+T_{e2}^2}$ is a real number, such that $R_{\pm}=1/2$ and $\Delta_{\pm}=-i\varepsilon_2\pm\alpha/2$. The two resonances are located at $\pm\alpha/2$ and have the same linewidth $\varepsilon_2$. The imaginary part of $\rho_{14}$ in the AT splitting regime can be written as
\begin{equation}
\mathrm{Im}(\rho_{14})_{\mathrm{ATS}}=-\left[\frac{\varepsilon_2/2}{(\Delta_p-\frac{\alpha}{2})^{2}+\varepsilon_2^2}+\frac{\varepsilon_2/2}{(\Delta_p+\frac{\alpha}{2})^{2}+\varepsilon_2^2}\right],
\end{equation}
where $\mathrm{Im}(\rho_{14})$ is determined by the sum of two identical Lorentzians peaked at $\pm\alpha/2$. Therefore, the overall absorption profile attributes to two Lorentzians, presenting a symmetric AT doublet in the limit of strong-tunneling regime.

Figure~\ref{SR}(b) indicates that, in the strong-tunneling regime, the reduction of absorption attributes to the contribution of two identical negative resonances which is termed as AT doublet~\cite{Cohen-Tannoudji1996}. It is interesting to see that the AT doublet is accompanied by double anticrossings in the eigenenergies of the system Hamiltonian [as shown in the red dashed loops in Fig.~\ref{eigenenergies1}(b)], forming a transparency window between the pair of resonances. The positive value of the resonance pair is only responsible for the decreasing or even vanishing absorption, instead of indicating a destructive nature of TIT. This is due to the fact that the pair of resonances is shifted away from each other, so that their overlap is insufficient to yield significant quantum interference.

An AT splitting pattern with a well-resolved doublet appears in the limit of strong-tunneling regime [see Fig.~\ref{SR}(c)]. There are prominent double anticrossings in the eigenenergies of the system Hamiltonian [as shown in the red dashed loops in Fig.~\ref{eigenenergies1}(c)]. This also results in an evident reduction of the overall absorption, displaying a wide transparency window (corresponding to vanishing absorption) when the tunneling couplings are sufficiently strong ($T_{e1}, T_{e2}\geq\Gamma_4$). In this case, the transparency window of the probe field is fully caused by the doublet structure rather than the quantum interference induced by TIT.

\section{Double Tunneling-induced transparency and Autler-Townes triplet}\label{s4}
\begin{figure}[htbp]
\begin{center}
\includegraphics[scale=0.8]{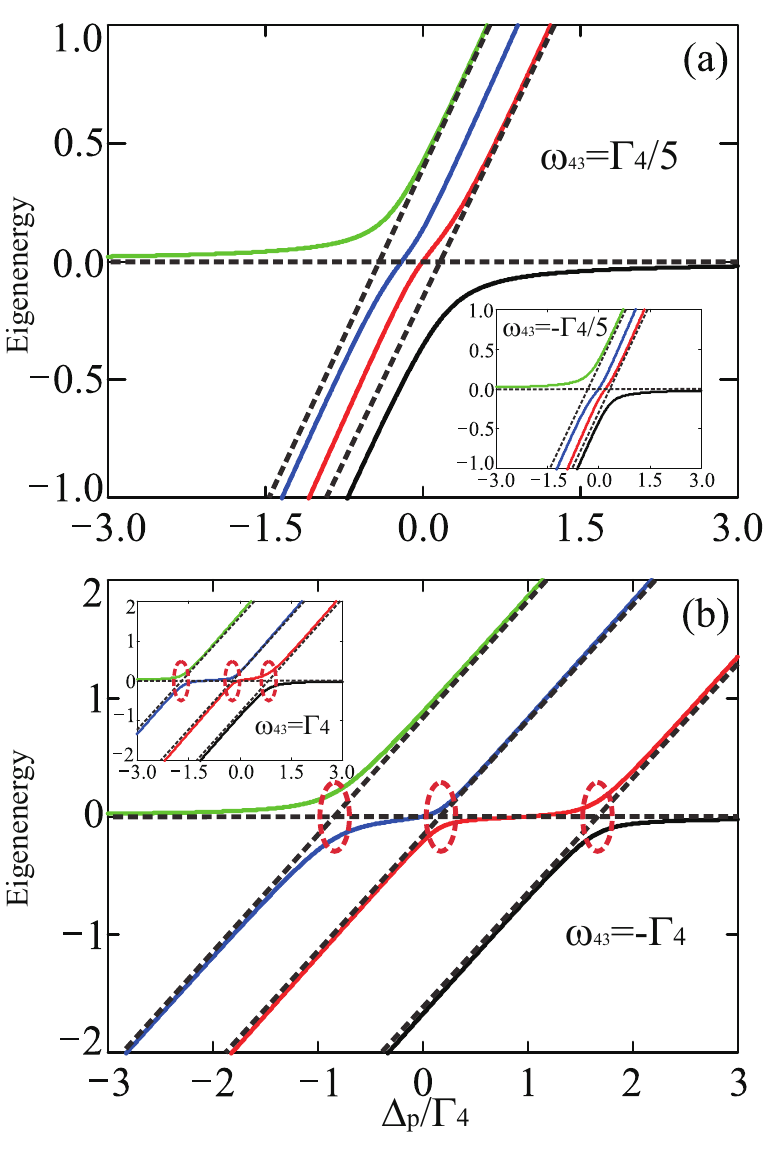}
\caption{(Color online) Eigenenergies of the Hamiltonian (\ref{H}) as a function of the probe-field detuning $\Delta_p$ at different values of the tunneling couplings strength and the frequency difference $\omega_{43}$: (a) $T_{e1}=T_{e2}=\Gamma_4/5$, $\omega_{43}=\pm\Gamma_4/5$, and (b) $T_{e2}=2T_{e1}=\Gamma_4$, $\omega_{43}=\pm\Gamma_4$, with $\omega_{42}=0$. Others parameters are the same as in Fig.~\ref{eigenenergies1}. The red dashed loops are used to highlight the anticrossing points.} \label{eigenenergies2}
\end{center}
\end{figure}
\begin{figure*}[htbp]
\begin{center}
\includegraphics[width=0.8\textwidth]{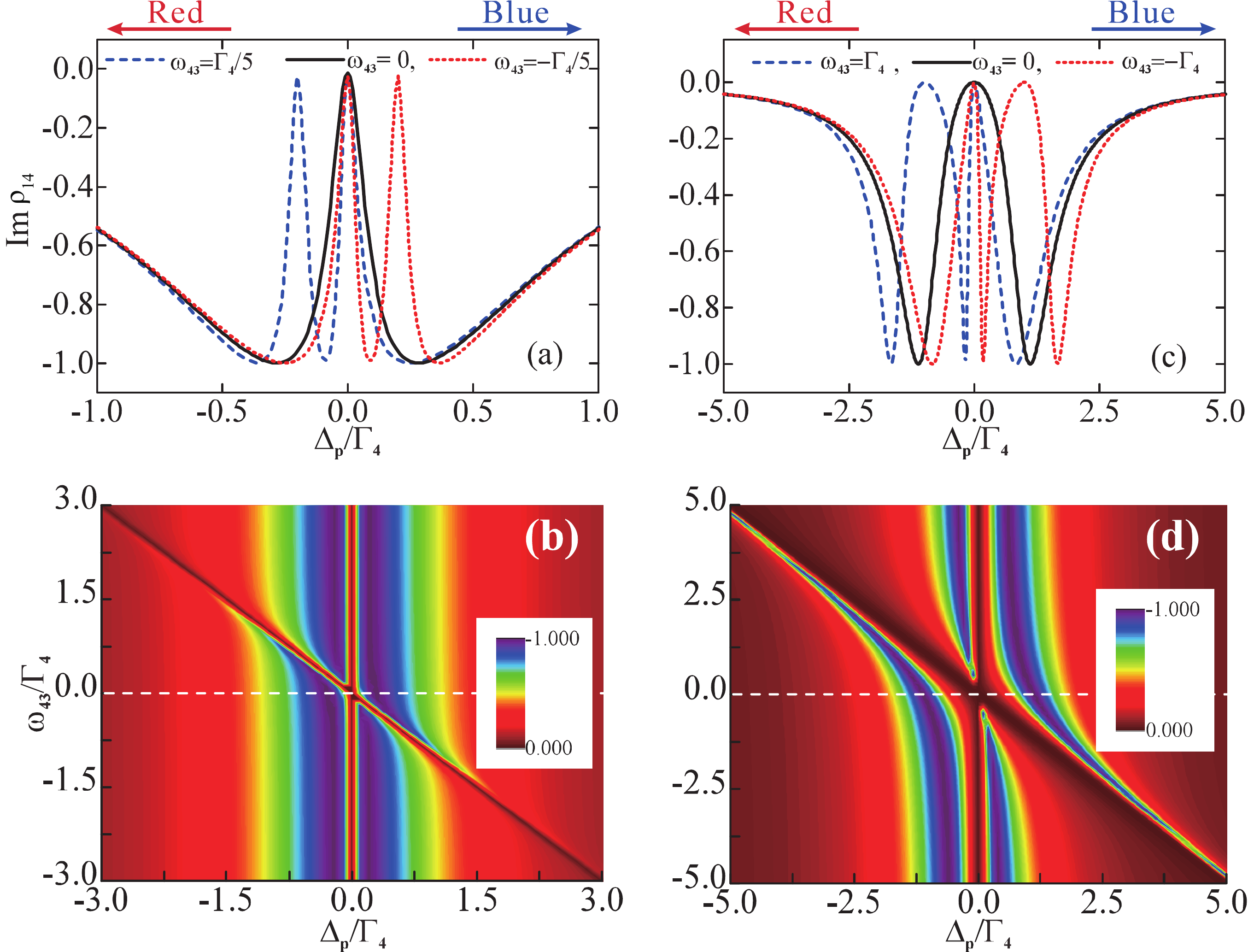}
\caption{(Color online) (a) and (b) $\mathrm{Im}(\rho_{14})$ as a function of the probe-field detuning $\Delta_p$ at different values of the frequency difference $\omega_{43}$, where $T_{e1}=T_{e2}=\Gamma_4/5$; (c) and (d) $\mathrm{Im}(\rho_{14})$ as a function of both the probe-field detuning $\Delta_p$ and the frequency difference $\omega_{43}$, where $T_{e2}=2T_{e1}=\Gamma_4$. Here $\omega_{42}=0$ and others parameters are the same as in Fig.~\ref{eigenenergies1}. } \label{DTITATT}
\end{center}
\end{figure*}

Next, we focus on the manipulation of the optical absorption of the probe field by varying $\omega_{43}$, which is a way of controlling the detuning $\Delta_3$. It should be noted that, in the weak-tunneling regime, the degenerate points of the eigenenergies of the system Hamiltonian disappear when $\omega_{43}$ is nonzero. For instance, when $\omega_{43}=\pm\Gamma_4/5$, no degenerate point is seen in Fig.~\ref{eigenenergies2}(a), which is different from the case in Fig.~\ref{eigenenergies1}(a). In the strong-tunneling regime, one can also see that the eigenenergies of the system Hamiltonian exhibit triple anticrossings [see Fig.~\ref{eigenenergies2}(b)], as compared to the case with double anticrossings in Fig.~\ref{eigenenergies1}(b) where $\omega_{43}=0$.

\subsection{The weak-tunneling regime with $\omega_{43}\neq0$}

As shown in Fig.~\ref{DTITATT}(a), in the weak-tunneling regime, it is revealed that double TIT can be realized by manipulating the energy-level detunning $\Delta_3$ to achieve slight off-resonance. Narrow double transparency windows arise, when the tunneling couplings are weeker than or equal to the threshold value $\mathrm{T_{t}^{'}}$, in the case of $\omega_{43}\neq0$. In particular, the new TIT dip [see the blue dashed curve in Fig.~\ref{DTITATT}(a)] is red-shifted for a blue-detuned $\Delta_3$ (e.g., $\omega_{43}=\Gamma_4/5$). However, the new TIT dip [see the red dotted curve in Fig.~\ref{DTITATT}(a)] becomes blue-shifted at a red-detuned $\Delta_3$ (e.g., $\omega_{43}=-\Gamma_4/5$). Meanwhile, the eigenenergies of the system Hamiltonian do not present the degenerate point and anticrossing [see Fig.~\ref{eigenenergies2}(a)], which also support the argument that the new TIT dip is detuning-dependent. Moreover, as shown in Fig.~\ref{DTITATT}(b), one of the absorption minima obeys the condition $\Delta_p=\omega_{42}=0$, and the other absorption minimum satisfies the condition $\Delta_p=-\omega_{43}$.  Therefore, we are able to realize double TIT without forming anticrossing in such a weak-tunneling regime.

\subsection{The strong-tunneling regime with $\omega_{43}\neq0$}

In the strong-tunneling regime, the transparency window of the AT doublet [see the black solid curve in Fig.~\ref{DTITATT}(c)] exhibits a new peak, turning the total absorption profile into three peaks. This yields two transparency windows [as shown in the blue dashed and red dotted curves in Fig.~\ref{DTITATT}(c)] termed as AT triplet. Interestingly, the wide transparency window is blue-shifted (red-shifted) at a red-detuned (blue-detunned) $\Delta_3$, e.g., $\omega_{43}=\mp\Gamma_4/5$. Also, one can see that there are triple anticrossings in the eigenenergies of the system Hamiltonian [see Fig.~\ref{eigenenergies2}(b)]. This behavior indicates that the wide transparency window is detuning-dependent.

In Fig.~\ref{DTITATT}(d), similar features can be observed, where two absorption minima locate at $\Delta_p=\omega_{42}=0$ and $\Delta_p=-\omega_{43}$, respectively. It should be noted that the width of the central peak increases when rasing the blue-detuning (red-detuning) $\Delta_3$, but the width of the peak on the red-detunned (blue-detunned) side decreases. In particular, for a blue-detunned (red-detunned) $\omega_{43}$, the width decrease of the red-detunned (blue-detunned) sideband is compensated by the width increase of the central peak. Therefore, in the strong-tunneling regime, the AT triplet can be realized by manipulating the detunning $\Delta_3$.

\section{Conclusion}\label{s5}

In summary, we have presented a theoretical study of the optical properties of a triple QD with four effective energy levels. The results show that, in the resonant-tunneling case of the triple QD, there is no anticrossing in the eigenenergies of the system Hamiltonian in the presence of TIT. This TIT leaves the total absorption profile of the probe field almost unchanged, but in the weak-tunneling regime a narrow transparency dip is observed. The collective contributions from the two excitation pathways, denoted as ``I'' and ``II'', lead to the cancellation of the overall absorption profile of the probe field due to destructive interference. However, in the strong-tunneling regime, the AT splitting exhibits two well resolved doublet and double anticrossings. For the off-resonance case, the double TIT gives rise to a new TIT dip in the weak-tunneling regime by manipulating one of the energy-level detuning in the absence of anticrossing. Moreover, in the presence of AT triplet, there are triple anticrossings in the eigenenergies of the system Hamiltonian. The emergence of the wide transparency window is red-shifted (blue-shifted) for a blue-detunned (red-detunned) level detuning. The linewidth narrowing in one of the side peaks compensates for the linewidth broadening in the central peak.

\begin{acknowledgments}
This work is supported by the National Natural Science Foundation of China Grant Nos.~91121015 and  11404019, the National Basic Research Program of China Grant Nos.~2014CB921401 and 2014CB848700, the NSAF Grant No.~U1330201,  and the China Postdoctoral Science Foundation Grant No.~2012M520146.
\end{acknowledgments}

\end{document}